%% file: main.tex
\documentclass{Interspeech}
\usepackage{multirow}
\usepackage{array} 
\usepackage{pifont}
\usepackage{lipsum}

\newcommand\blfootnote[1]{%
\begingroup
\renewcommand\thefootnote{}\footnote{#1}%
\addtocounter{footnote}{-1}%
\endgroup
}

\interspeechcameraready

\title{Mel-McNet: A Mel-Scale Framework for Online Multichannel Speech Enhancement}

\author[affiliation={1,2}]{Yujie}{Yang}
\author[affiliation={2,3}]{Bing}{Yang}
\author[affiliation={2,3}]{Xiaofei}{Li$^{*,}$} 

\affiliation{}{Zhejiang University}{China}
\affiliation{School of Engineering}{Westlake University}{China}
\affiliation{Institute of Advanced Technology}{Westlake Institute for Advanced Study}{China}
\email{yangyujie@westlake.edu.cn, yangbing@westlake.edu.cn, 
lixiaofei@westlake.edu.cn}
\keywords{multichannel speech enhancement, real-time, Mel-scale, nonlinear-frequency scale}

\usepackage{comment}

\begin{document}

\maketitle

\begin{abstract}
 Online multichannel speech enhancement has been intensively studied recently. Though Mel-scale frequency is more matched with human auditory perception and computationally efficient than linear frequency, few works are implemented in a Mel-frequency domain. To this end, this work proposes a Mel-scale framework (namely Mel-McNet). It processes spectral and spatial information with two key components: an effective STFT-to-Mel module compressing multi-channel STFT features into Mel-frequency representations, and a modified McNet backbone directly operating in the Mel domain to generate enhanced LogMel spectra. The spectra can be directly fed to vocoders for waveform reconstruction or ASR systems for transcription. Experiments on CHiME-3 show that  Mel-McNet can reduce computational complexity by 60\% while maintaining comparable enhancement and ASR performance to the original McNet. Mel-McNet also outperforms other SOTA methods, verifying the potential of Mel-scale speech enhancement.

\end{abstract}
\vspace{-0.1cm}
\blfootnote{* Corresponding author}
\input{sections/1_Introduction}

\vspace{-0.1cm}
\input{sections/2_Method}
\vspace{-0.1cm}

\input{sections/0_table_onlineSE_performance}

\input{sections/0_table_abation_study_new}
\input{sections/3_Experiments}
\vspace{-0.2cm}
\input{sections/4_Conclusion}
\newpage

\bibliographystyle{IEEEtran}
\bibliography{mybib}

\end{document}

%% file: sections/1_Introduction.tex
\section{Introduction}
Speech enhancement aims to separate target speech from background noise. It has gained increasing importance in emerging smart devices, such as smart homes and embodied intelligence systems. These applications pose high requirements on not only real-time implementation and low computational complexity for deployment on edge devices, but also a good performance on both speech quality and intelligibility.

Recently, deep learning-based approaches have shown superiority for multichannel speech enhancement \cite{SE_overview_wang}, which can be broadly categorized into neural beamforming methods and end-to-end methods. Neural beamforming methods employ deep neural networks (DNNs) to estimate parameters of the beamformer, such as the steering vector of the target speech signal, and the spatial covariance matrices (SCMs) of target speech and undesired noise signals \cite{MNMF_beamforming_Shimada,blstm_gev_beamformer_Heymann}. They are typically performed in narrow bands, which leads to high computation costs. 
In contrast, end-to-end methods directly estimate target speech signals using DNNs. Among them, the combination of full-band and narrow-band (or sub-band) processing \cite{spatialnet_quan,tfgridnet_wang,ftjnf_tesch,ospatialnet_quan} shows significant superiority. McNet proposed in \cite{mcnet} utilizes four dedicated and cascaded modules to learn full-band spatial, narrow-band spatial, sub-band spectral, and full-band spectral information respectively. The online version of the McNet shows superior speech enhancement performance. However, it suffers from high computation costs due to the use of stacked along-frequency and along-time long short-term memory (LSTM) blocks.

We find that most existing works are implemented in the time domain \cite{fasnet_luo,tparn_pandey} or in the linear-frequency scale time-frequency (T-F) domain (e.g., the short-time Fourier transform (STFT) domain) \cite{spatialnet_quan,tfgridnet_wang,ftjnf_tesch,multichannel_SEwithoutbeamforming}. Only a few approaches \cite{melfullsubnet,melsubband_beamformer_tecent,gammatone_tasnet_timo} explore the nonlinear/perception-aware frequency scale, such as the Mel scale, Gammatone scale, and Equivalent Rectangular Bandwidth (ERB) scale. 
Nonlinear frequency scale has compact and perceptually aligned frequency resolution, namely higher resolution at lower frequencies and lower resolution at higher frequencies. Compared with linear frequency, this scale setting mainly provides two benefits: it better conforms to human speech perception of the spectral envelope and signal periodicity, and it offers lower computational complexity, particularly for narrow-band (or sub-band) methods. In addition, most automatic speech recognition (ASR) systems utilize logarithmic Mel (LogMel) spectrograms as standard input features, which demonstrates the potential of the Mel domain processing for effective information extraction. These observations naturally motivate us to \textbf{directly enhance speech in a nonlinear-frequency scale T-F domain.} 

Some works have attempted to process speech signals in a nonlinear-frequency scale. Zhou et al.~\cite{melfullsubnet} explored the spectral information in the Mel domain for single-channel speech enhancement, which directly inputs the LogMel spectrogram into a DNN and outputs the enhanced LogMel spectrogram. For multichannel speech enhancement \cite{spatialnet_quan,ftjnf_tesch,mcnet,eabnet_li}, both spectral and spatial information are important for distinguishing speech from noise. Kothapally et al.~\cite{melsubband_beamformer_tecent} proposed a Mel-scale subband beamformer for speech separation. It compresses the full-band SCMs into Mel-scale representation for further subband model processing, and then recovers to the full-band scale to obtain separated signals. Though Mel-scale compression can reduce computation complexity to some degree, the full-band-scale recovery seems redundant for Mel-domain ASR backends. How to effectively use both spectral and spatial information in a nonlinear frequency scale remains a challenging problem. 

\input{sections/0_diagram_melmcnet}
This work proposes a novel online multichannel speech enhancement framework in the Mel domain, termed Mel-McNet.
Compared with the original McNet, the proposed Mel-McNet compresses multichannel STFT features into Mel-scale representations via an effective STFT-to-Mel module. It processes the spectral and spatial information directly in the Mel domain with a modified McNet backbone. The enhanced LogMel Spectra generated by the Mel-McNet could be either fed to a vocoder for waveform reconstruction or an ASR system for transcription. 
Experiments show the proposed Mel-McNet could reduce computational complexity by nearly 60\% while still maintaining competitive performance for both speech quality and intelligibility, which reveals the feasibility and potential of Mel-scale multichannel speech enhancement. Code is available at \footnote{https://github.com/Audio-WestlakeU/Mel-McNet.git}.

%% file: sections/0_diagram_melmcnet.tex
\begin{figure*}
    \centering
    \includegraphics[width=0.94\textwidth]{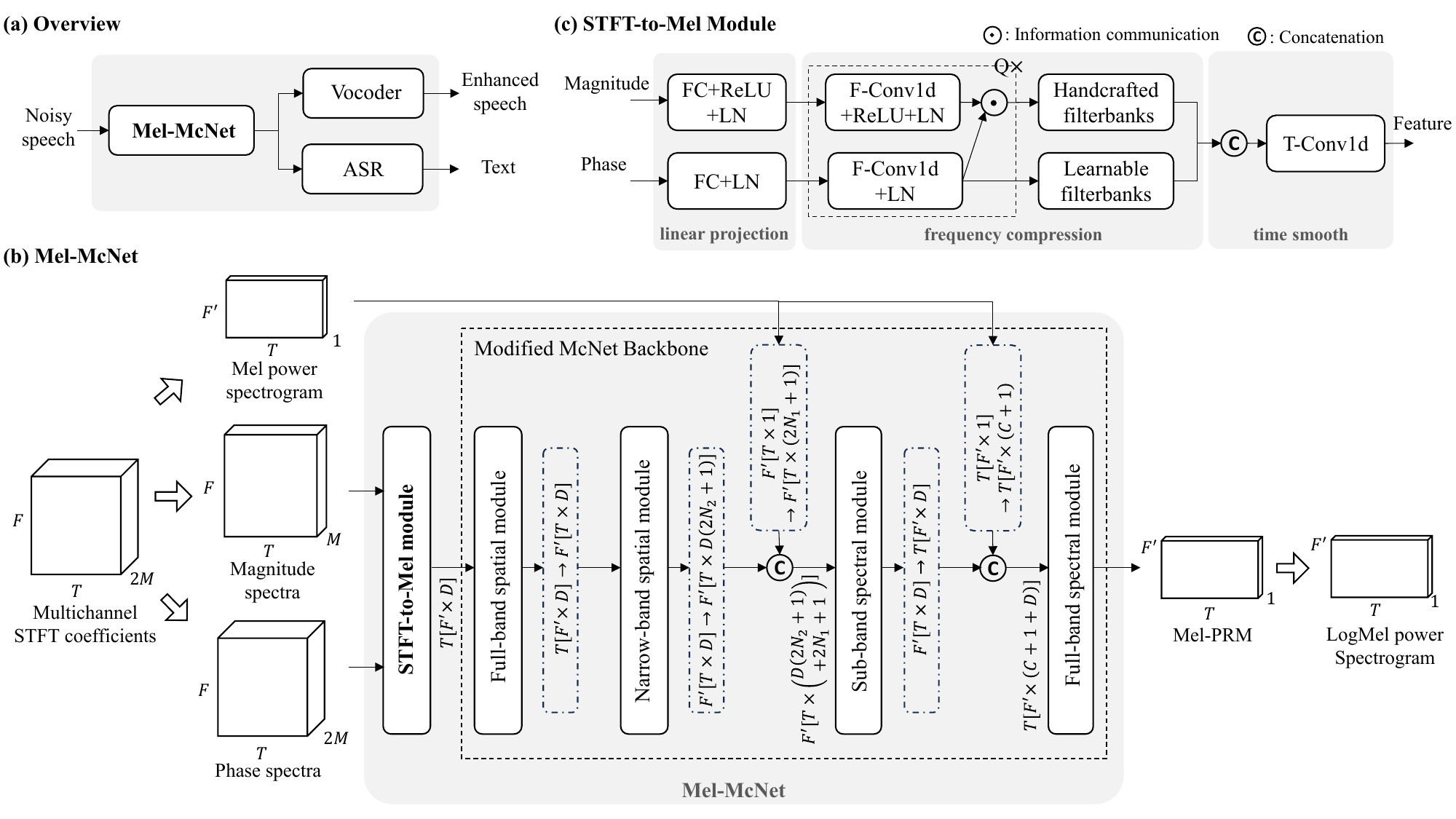}
    \vspace{-0.2cm}
    \caption{The proposed Mel-scale framework. (a) The system overview. (b) The diagram of Mel-McNet. (c) The architecture of STFT-to-Mel module. $F$, $F'$, $T$, and $M$ represent the number of linear-scale STFT frequencies, nonlinear Mel-scale frequencies, time frames, and microphones, respectively. $N_{1}$ and $N_{2}$ are the number of adjacent frequencies. $C$ is the number of context frames. $D$ is the dimension of hidden embeddings. For the McNet backbone, the feature dimension follows the form of "batch dimension[dimension of one sample in a batch]", and the dash-dotted boxes indicate dimension transformation operations (please see details in \cite{mcnet}). }
    \label{fig:diagram_of_melmcnet}
    \vspace{-0.5cm}
\end{figure*}


%% file: sections/2_Method.tex
\section{Method}
The proposed Mel-scale framework for multichannel speech enhancement is illustrated in Fig.\ref{fig:diagram_of_melmcnet} (a). The noisy speech is fed into Mel-McNet to obtain the enhanced LogMel spectrogram, which is followed by a vocoder to reconstruct the speech signal in the time domain and/or an ASR model for text transcription. 
\vspace{-0.1cm}
\subsection{Mel-McNet}
As depicted in Fig.\ref{fig:diagram_of_melmcnet}(b), the proposed Mel-McNet is derived from McNet \cite{mcnet}. To facilitate the backbone ASR application and meanwhile guarantee a lightweight speech enhancement model, a STFT-to-Mel module (see details below) is added before the McNet, which converts the linear frequency into Mel subband and hence reduces the following frequency-wise computation complexity of McNet. The Mel-McNet utilizes multichannel magnitude and phase spectra in the STFT domain as input, and predicts the Mel-scale power ratio mask (Mel-PRM) of the reference channel. When training the Mel-McNet, the mean square error (MSE) between the predicted and target mask is utilized as the loss function. During inference, to obtain the enhanced LogMel power spectrogram, the square of the ratio mask is multiplied with the power spectrogram of the noisy speech, and then taken logarithm operation. 
\vspace{-0.1cm}
\subsubsection{STFT-to-Mel module} 
This module compresses STFT-domain multichannel magnitude and phase features into Mel-domain representations. As shown in Fig. \ref{fig:diagram_of_melmcnet}(c), the magnitude and phase spectrograms are successively passed to a linear block, a frequency compression block (FCB) and a time smooth block (TSB). The linear block is a fully connected (FC) layer followed by (a ReLU activation and) a layer normalization (LN) layer. 

The FCB block consists of $Q$ one-dimensional convolutional layers (F-Conv1d) with (ReLU and) LN, and handcrafted/learnable filterbanks, which can capture and compress the local frequency patterns. The F-Conv1d is with a kernel size along the frequency axis and $D$ channels. 
Spectral and spatial cues have different properties along frequency. Speech content (spectral cues) mainly exists at lower frequencies and the Mel scale better matches human auditory perception, while inter-channel phase differences (spatial cues) are linearly proportional to frequency. Accordingly, though the multichannel magnitude and phase features are processed along the same pipeline, the magnitude branch is with nonlinear or special operations including ReLU activation functions and handcrafted Mel-scale filterbanks, while the phase branch is without ReLU and uses learnable filterbanks. 
After passing each convolution block, we obtain two hidden representations $\bold{E}^{\rm Mag}_{i} \in \mathbb{R}^{T \times F \times D}$ and $\bold{E}^{\rm Pha}_{i} \in \mathbb{R}^{T \times F \times D}$, where $i$ represents the index of the along-frequency convolution blocks. An attention-like information communication operation (denoted as $\odot$) \cite{PHASEN_Luo2020} is used to exploit the mutual information between these two types of features, which is defined as: 
\setlength{\abovedisplayskip}{8pt}
\setlength{\belowdisplayskip}{8pt}
\begin{align}
    \bold{E}^{\rm Mag}_{i} \odot \bold{E}^{\rm Pha}_{i} = \bold{E}^{\rm Mag}_{i} \circ tanh(linear(\bold{E}^{\rm Pha}_{i})) \notag,
\end{align}
where $\circ$ denotes the element-wise multiplication, and $linear(\cdot)$ represents a linear layer.
The triangular filters \cite{Mel_Stevens} is taken as the handcrafted Mel-scale filterbanks, while a simple fully connected layer is employed as the learnable filterbanks. Both transform feature dimension from $T \times F \times D$ to $T \times F^{'} \times D$.

Finally, the two-branch features are stacked along the embedding dimension, obtaining a feature with a dimension of $T \times F^{'} \times 2D$. Then the stacked embedding is fed into a one-dimension temporally causal convolution layer to smooth these hidden representations over time. The output of the STFT-to-Mel module has a dimension of  $T \times F^{'} \times D$. 
\vspace{-0.1cm}
\subsubsection{Learning target}
We adopt a rectified Mel-PRM of the reference channel as the learning target. For each Mel-scale T-F bin, it is formulated as:
\setlength{\abovedisplayskip}{3pt}
\setlength{\belowdisplayskip}{3pt}
\begin{align}
    \text{Mel-PRM} = {\rm min}(\sqrt{\frac{S^{\rm Mel}_{r}(t,f^{'})}{X^{\rm Mel}_{r}(t,f^{'})}}, 1) \notag,
\end{align}
where $f^{'}\in[0, F^{'}-1]$ denotes the Mel-scale frequency index and $r$ denotes the index of reference microphone. $S^{\rm Mel}_{r}(t,f^{'})$ and $X^{\rm Mel}_{r}(t,f^{'})$ are the Mel-scale power of clean speech and noisy speech, respectively. The Mel-scale power spectrogram is computed by filtering the linear-scale frequencies with triangular filterbanks \cite{Mel_Stevens}. 


\vspace{-0.1cm}
\subsubsection{McNet backbone}
The Mel-scale multichannel features output by the STFT-to-Mel module is passed into the McNet backbone \cite{mcnet}. McNet consists of four cascaded modules to respectively process the full-band spatial, narrow-band spatial, sub-band spectral, and full-band spectral information. The first two modules exploit multichannel spatial information, while the last two exploit single-channel spectral information. Each module is composed of a (along-frequency/along-time) (Bi)LSTM and a linear layer. 

Minimal changes are made to the original McNet backbone: (a) The sub-band spectral and full-band spectral modules take the Mel-domain power spectrogram of the reference channel as the supplementary information; (b) It uses the Sigmoid activation function at the last layer to guarantee that the output value ranges from 0 to 1. 
\vspace{-0.1cm}
\subsection{Backends}
During inference, the proposed Mel-McNet can be followed by either a vocoder to obtain the enhanced speech signal in the time domain, or an ASR model to get the text transcription.

\textbf{Vocoder}: 
Similar to the work \cite{melfullsubnet}, we use the neural vocoder to transform the enhanced LogMel spectrogram to the time-domain waveform. The Vocos \cite{vocos_siuzdak}, a well-performed and effective neural vocoder based on generative adversarial networks (GANs), is adopted in our work.

 
\textbf{ASR}: 
The enhanced LogMel spectrogram can be directly fed into a pre-trained ASR system without fine-tuning or joint training. To evaluate the speech recognition performance of enhancement methods, we use a pre-trained offline ASR model based on E-branchformer provided in the ESPNet toolkit \footnote{https://github.com/espnet/espnet/tree/master/egs2/chime4/asr1}.


%% file: sections/0_table_onlineSE_performance.tex
\begin{table*}[tbp]
\caption{Online multichannel speech enhancement results on 6-channel simulated CHiME-3 dataset. For Mel-McNet, the Param. and FLOPs of speech enhancement network/neural vocoder are given in addition.}
\vspace{-0.2cm}
\label{tab:comparsion_with_other_methods}
\centering
\resizebox{\linewidth}{!}{ 
\begin{tabular}{l|c|c|ccc|c|c|c}
\toprule
\multirow{2}{*}{Method} & \multirow{2}{*}{Param.(M)} & \multirow{2}{*}{FLOPs(G/s)} & \multicolumn{3}{c|}{DNSMOS(P.835)} & \multirow{2}{*}{WB-PESQ $\uparrow$} & \multirow{2}{*}{STOI (\%) $\uparrow$} & \multirow{2}{*}{WER (\%) $\downarrow$} \\
& & & MOS-SIG $\uparrow$ & MOS-BAK $\uparrow$ & MOS-OVR $\uparrow$ & & & \\
\midrule
Noisy speech& - & - & 3.11 & 1.90 & 1.97 & 1.27 & 87.0 & 9.1 \\
\midrule
Clean speech& - & - & 3.55 & 4.06 & 3.28 & 4.50 & 100.0 & 3.1 \\
Clean Mel+Vocos & - & - & 3.55 & 4.05 & 3.28 & 3.63 & 97.7 & - \\
\midrule
EaBNet \cite{eabnet_li}& 3.27 & 17.97 & 3.40 & 3.90 & 3.08 & 2.46 & 97.1 & 5.6\\
oSpatialnet-mamba \cite{ospatialnet_quan}& 1.74 & 71.91 & 3.39 & 3.95 & 3.09 & 2.63 & 97.4 & 5.4 \\
McNet \cite{mcnet}&  1.85 & 115.13 & 3.40 & \textbf{3.99} & 3.12 & \textbf{2.80} & \textbf{97.7} & \textbf{4.4} \\
\midrule
Mel-McNet (proposed) & 15.24(2.04+13.20)& 46.96(43.75+3.21) &  \textbf{3.43} & \textbf{3.99} & \textbf{3.14} & 2.44 & 95.4 & \textbf{4.4} \\
\bottomrule
\end{tabular}}
\vspace{-0.4cm}
\end{table*}

%% file: sections/0_table_abation_study_new.tex
\begin{table}[tbp]
\setlength\tabcolsep{4pt}
\renewcommand{\arraystretch}{1.4}
\caption{Ablation study of STFT-to-Mel module. }
\vspace{-0.2cm}
\label{tab:ablation_study_new}
\centering
\resizebox{\linewidth}{!}{ 
\begin{tabular}{c|c|ccc|c|c}
\toprule
 Feature & \multicolumn{3}{c|}{Compression Module} & MOS-OVR $\uparrow$ & WB-PESQ $\uparrow$ & WER (\%) $\downarrow$ \\
\midrule
 \multirow{5}{*}{Joint} & \multicolumn{3}{c|}{TrainMel \cite{trainMel_tecent}} & 3.08 & 2.27 & 5.1\\
 & \multicolumn{3}{c|}{HandcraftedMel} & 3.07 & 2.36 &4.7\\
 & \multicolumn{3}{c|}{FCB} & 3.12 & 2.41 & 4.6 \\
 & \multicolumn{3}{c|}{FCB+TSB} & 3.11 & 2.39 & 4.5\\
\midrule
Separate & \multicolumn{3}{c|}{FCB+TSB (proposed)} & \textbf{3.14} & \textbf{2.44} & \textbf{4.4} \\

\bottomrule
\end{tabular}}
\vspace{-0.4cm}
\end{table}

%% file: sections/3_Experiments.tex
\section{Experiments}
\vspace{-0.1cm}
\subsection{Experimental Setup}
\textbf{Dataset}: 
We evaluate both speech enhancement and recognition performance on the official evaluation and test sets (comprising 1,640 and 1,320 utterances, respectively) of the simulated CHiME-3 dataset \cite{chime3}. 

The training set of speech enhancement model is constructed by combining clean speech signals from the DNS Challenge \cite{dns5_dubey} with noise signals from the CHiME-3 dataset \cite{chime3}. Specifically, we select 100-hour high-quality clean speech as the source signal, achieving an average raw MOS OVERALL (P.835) score of 4.18. It consists of a total of 1,668 speakers. The official CHiME-3 script is used to generate reverberation-free 6-channel clean speech signals (only the time delay is considered). The multichannel background noise signals are recorded across four environments, namely bus, café, pedestrian area, and street junction, using a tablet device equipped with six microphones. The recordings are sampled at 16 kHz. To generate the training data, noise clips are randomly selected and mixed with clean speech according to a signal-to-noise ratio (SNR) uniformly ranging from -5 dB to 10 dB. The data is generated on-the-fly during training.

For the training of vocoder, the clean speech signals are collected from six datasets, including AISHELL I \cite{aishell1_bu}, AISHELL II \cite{aishell2_du} and THCHS30 \cite{thchs30_wang} for Chinese, and EARS \cite{ears_richter}, VCTK \cite{VCTK_Yamagishi} and DNS challenge \cite{dns5_dubey} for English. About 200-hour high-quality speech data are selected for each language.

The ASR model is trained on the mixture of "tr05" and "si284" of the Wall Street Journal (WSJ) corpus \cite{wsj_paul92_icslp} dataset as provided in the ESPNet toolkit.

\textbf{Parameter Settings}: 
The sample rate is 16kHz. For the proposed Mel-Mcnet, the Vocos and the ASR model, STFT is all performed using a 512-sample (32ms) Hanning window with a frame step of 128 (8ms) samples. The number of Mel frequencies $F'$ is set to 80. The noisy multichannel signals are normalized in the STFT domain before being processed by the network. The normalization scheme follows the work \cite{mcnet}. 

\textbf{Model Configurations}: 
For Mel-Mcnet, the dimension of hidden embedding dimensions $D = 64$. The number of F-Conv1d blocks $Q$ is 3 and the kernel size is 3. As for the 1-layer T-Conv1d, the kernel size is 6 and we pad the past 5 frames with zeros to make it temporally causal. The configurations of McNet backbone is the same as that in \cite{mcnet}.  
We select the fifth microphone as the reference channel.

For Vocos, we keep almost the same configurations as its original paper, except two minor changes to fit the Mel-Mcnet. One is replacing magnitude-based Mel spectrograms with power-based ones, and preliminary experiments show that this change does not affect the performance of Vocos. The other is converting original offline Vocos to online processing by replacing non-causal convolutions with causal ones, and competitive performance can be achieved after this change \cite{melfullsubnet}. 

\textbf{Training Details}: Adam is used as the optimizer with a learning rate of 0.001. The batch size is set to 4. We train our model until convergence, which takes almost 30 epochs.

\textbf{Comparison Methods}: 
We compare the proposed method with the following online multichannel speech enhancement methods.
1) EaBNet \footnote{https://github.com/Andong-Li-speech/EaBNet} \cite{eabnet_li} is an advanced causal and framewise neural beamformer with two core modules to learn 3-D spatial-spectral embedding tensor and estimate the filter weights. 
2) oSpatialnet-mamba \footnote{https://github.com/Audio-WestlakeU/NBSS} \cite{ospatialnet_quan}, the online extension of the Spatialnet \cite{spatialnet_quan}, is an outstanding end-to-end neural spatial filter, which has a strong ability to exploit the spatial information for discriminating between the target speech and noise.
3) McNet \footnote{https://github.com/Audio-WestlakeU/McNet} \cite{mcnet}, the backbone of the proposed method.

For a fair comparison, all comparison methods are re-trained on the training set mentioned above, and  share the same STFT configurations in order to have the same resolution along both the time and frequency axes of the input features. Except that, we keep the same configurations as their original settings. 

\textbf{Evaluation Metrics}: 
Both instructive (namely wideband perceptual evaluation of speech quality (WB-PESQ) \cite{pesq} and short-time objective intelligibility (STOI) \cite{stoi}) and non-instructive metrics (namely DNSMOS P.835 \cite{dnsmos835}) are employed to evaluate speech enhancement performance. 
The word error rate (WER) is adopted to evaluate ASR performance.

\vspace{-0.1cm}
\subsection{Results and Analysis} 
\textbf{Upper-Bound Performance}: 
To better guide the following experiments, we measure the upper-bound performances of two frameworks, i.e., \emph{"clean speech"} and \emph{"clean Mel+Vocos"}. 
The former refers to the clean speech of the reference channel in the time domain, while the latter involves extracting the LogMel power spectrogram from the same clean speech and subsequently converting it back to the time domain using a pre-trained vocoder. 
It can be seen in Table \ref{tab:comparsion_with_other_methods} that instructive metrics such as WB-PESQ and STOI, the \emph{"clean Mel+Vocos"} yields lower scores than \emph{"clean speech"}. In contrast, for non-instructive metrics like DNS-MOS (P.835), no significant performance difference is observed between these two. This performance discrepancy is mainly caused by the pre-trained vocoder which transforms the LogMel power spectrogram to the time-domain signal. The vocoder can generate perceptually high-quality speech, but may not be very numerically close to the original signal.  Consequently, in the following, we provide more references to non-instructive metrics and WER, and WB-PESQ and STOI are provided for supplementary reference.

\textbf{Comparison with State-of-The-Art (SOTA) Methods}: 
The proposed framework introduces a moderate increase in parameters relative to McNet (attributed to the Vocos vocoder) but achieves a 60\% reduction in FLOPs \footnote{We use the official tool provided by PyTorch (torch.utils.flop counter.FlopCounterMode) for FLOPs computation.}. The FLOPs reduction is mainly attributed to data processing in the compressed Mel frequency domain.  
The proposed method achieves SOTA results on the CHiME-3 benchmark in terms of DNSMOS and WER, which demonstrates the proposed method can generate perceptually higher-quality speech than other methods. It obtains slightly worse WB-PESQ and STOI mainly because the vocoder can not guarantee a close value to the original signal (as analyzed in \textbf{Upper-bound performance}).
Although EaBNet has lower computational complexity, the ASR performance is heavily impacted. 
We evaluate the real-time factor (RTF) of the proposed enhancement model on a platform equipped with an Intel(R) Core(TM) i7-9700 CPU @ 3.00GHz. The overall RTF is approximately 0.799, with Mel-McNet and Vocos contributing 0.755 and 0.044, respectively. These experiments guarantee real-time deployment of our method in real-world applications without compromising performance. 

\textbf{Ablation Study}: This study compares different STFT-to-Mel compression modules within our framework. Table \ref{tab:ablation_study_new} presents the experimental results. In this table, the terms "Joint" and "Separate" denote two ways of organizing multichannel input features. "Joint" represents stacking the real and imaginary components of multichannel STFT coefficients and taking them as a whole, while "Separate" denotes feeding the neural network with separate multichannel magnitude and phase spectra.  
"TrainMel" \cite{trainMel_tecent} uses an $F'$-group fully connected layer to compress frequency dimension, with Mel subband frequencies matching handcrafted Mel filterbanks. 
HandcraftedMel uses the Handcrafted Mel-scale filterbanks, and FCB combines convolution blocks to it (see Fig.\ref{fig:diagram_of_melmcnet}(c)). 

For the joint feature setting, FCB performs better than HandcraftedMel and TrainMel, which owes to using $Q$ F-Conv1d to capture multi-resolution frequency patterns before Mel domain compression. The performance improvement from FCB to FCB+TSB indicates that time-axis smoothing improves speech recognition performance. 
When changing joint setting to separate setting, both speech enhancement and recognition performance are further improved, because spectral and phase spectra are separately processed in dual branches. 
The proposed STFT-to-Mel module achieves the best performance among all alternative compression schemes. 

%% file: sections/4_Conclusion.tex
\section{Conclusion}
This work proposes a Mel-scale framework for online multichannel speech enhancement. With an effective STFT-to-Mel module inserted before the McNet backbone, Mel-McNet can process both spectral and spatial information in the Mel domain. The enhanced LogMel spectrogram output by Mel-McNet can be passed to a vocoder/ASR model for reconstructing the enhanced speech signal in the time domain or text transcription. Compared with the backbone, the proposed framework maintains comparable performance of speech enhancement and recognition, but reduces the computational complexity by nearly 60\%. It also outperforms other SOTA online methods on the CHiME-3 dataset. This work reveals the feasibility and potential of multichannel speech enhancement at the Mel-scale. Future work will test the generalization of Mel-scale multichannel speech enhancement on other lightweight backbones.